\documentclass[11pt,a4paper]{article}

\usepackage{amsmath,amsfonts,graphicx,natbib,psfrag,color}
\usepackage{algorithm}
\usepackage[textsize=small]{todonotes}
\usepackage{booktabs}

\renewcommand{\vec}[1]{\boldsymbol{#1}}

\title{Efficient sampling of conditioned Markov jump processes}

\author{Andrew Golightly$^1$\footnote{andrew.golightly@ncl.ac.uk}, Chris Sherlock$^2$}
\date{\small $^1$ School of Mathematics, Statistics and Physics, Newcastle University, UK\\$^{2}$Department of Mathematics and Statistics, Lancaster University, UK\\}

\begin{document}
\maketitle
\begin{abstract}
We consider the task of generating draws from a Markov jump process (MJP) between two time-points at which the process is known. 
Resulting draws are typically termed \emph{bridges} and the generation of such bridges plays a key role in simulation-based 
inference algorithms for MJPs. The problem is challenging due to the intractability of the conditioned process, necessitating the 
use of computationally intensive methods such as weighted resampling or Markov chain Monte Carlo. An efficient implementation 
of such schemes requires an approximation of the intractable conditioned hazard/propensity function that is both cheap and accurate. 
In this paper, we review some existing approaches to this problem before outlining our novel contribution. Essentially, we leverage 
the tractability of a Gaussian approximation of the MJP and suggest a computationally efficient implementation of the resulting conditioned 
hazard approximation. We compare and contrast our approach with existing methods using three examples.
\end{abstract}


\section{Introduction}
\label{sec:intro}

Markov jump processes (MJPs) can be used to model a wide range of discrete-valued, 
continuous-time processes. {Our focus here is on the MJP representation 
of a reaction network, which has been ubiquitously applied} in areas such as 
epidemiology \citep{Fuchs_2013,lin2013b,mckinley2014}, population ecology \citep{matis07,BWK08} and systems biology 
\citep{Wilkinson09,Wilkinson06,sherlock2014}.  Whilst exact, forward simulation of {this class of} MJP is 
straightforward \citep{Gillespie77}, the reverse problem of performing fully Bayesian inference for the parameters governing 
the MJP given partial and/or noisy observations is made challenging by the intractability of the observed data likelihood. Simulation-based 
approaches to inference typically involve ``filling in'' event times
and types between the observation times. A key repeated step in many
inference mechanisms starts with a sample of possible states at one observation
time and for each element of the sample, creates a trajectory
starting with the sample value and ending at the time of the next
observation with a value that is consistent with the next observation.
The resulting conditioned samples are typically referred to as
\emph{bridges}, and ideally the bridge should be a draw from the exact
distribution of the path given the initial condition and the observation. 
However, except for a few simple cases, exact simulation of MJP bridges is infeasible, necessitating approximate 
bridge constructs that can be used as a proposal mechanism inside a weighted resampling and/or Markov chain Monte Carlo (MCMC) 
scheme. 

The focus of this paper is the development of an approximate bridge
construct that is both accurate and computationally efficient. Our
contribution  can be applied in a generic observation regime that
allows for discrete, partial and noisy 
measurements of the MJP, and is particularly
effective compared to competitors in the most difficult regime where
the observations are sparse in time and the observation variance is small. 
Many bridge constructs have been proposed for partially observed
stochastic differential equations (SDEs, e.g. \cite{DelHu06}, \cite{bladt14}, \cite{bladt16}, \cite{schauer17} and \cite{whitaker2017}) 
but the literature on bridges for MJPs is relatively sparse. Recent progress involves an approximation of the instantaneous rate or hazard function governing the 
conditioned process. For example, \cite{BWK08} linearly interpolate the hazard between observation times but require full 
and error-free observation of the system of interest. \cite{fearnhead2008} recognises that the conditioned hazard requires 
the intractable transition probability mass function of the MJP. This is then directly approximated by substituting 
the transition density associated with {the coarsest possible discretisation
  of a spatially }continuous approximation of the 
MJP, the chemical Langevin equation \citep{Gillespie00}. \cite{GoliWilk15} derive a conditioned hazard 
by approximating the expected number of events between observations, given the observations themselves. 
Unfortunately, the latter two approaches typically perform poorly when the behaviour of the conditioned process 
is nonlinear. 

We take the approach of \cite{fearnhead2008} as a starting point and replace the intractable MJP transition probability 
with the transition density governing the linear noise approximation (LNA) \citep{kurtz1970,elf2003,Komorowski09,schnoerr17}. 
{Whilst the LNA has been used as an inferential model (see e.g. \cite{ruttor09} and \cite{ruttor10} for 
a maximum likelihood approach and \cite{stathopoulos13} and
{\cite{fearnhead14}} for an MCMC approach), we believe 
that this is the first attempt to use the LNA to develop a bridge construct for simulation of conditioned MJPs.} We 
find that the LNA offers superior accuracy over a single step of the CLE (which must be discretised in practice), 
at the expense of computational efficiency. Notably, the LNA solution
requires, for each event time in each trajectory, integrating forwards until the next event time a system of ordinary 
differential equations (ODEs) whose dimension is quadratic in the number of MJP components. We therefore leverage 
the linear Gaussian structure of the LNA to derive a bridge construct that
only requires a single full integration of 
the LNA ODEs, irrespective of the number of {transition} events on each bridge or
the number of bridges required. We compare the resulting novel construct to several 
existing approaches using three examples of increasing complexity. In the final, real-data application, we demonstrate use of 
the construct within a pseudo-marginal Metropolis-Hastings scheme, for performing fully Bayesian inference 
for the parameters governing an epidemic model.

The remainder of this paper is organised as follows. In Section~\ref{sec:stochkin} we define a Markov jump process 
as a probabilistic description of a reaction network. We consider the task of sampling conditioned jump 
processes in Section~\ref{sec:cond}, and review two existing approaches. Our novel contribution is presented in 
Section~\ref{sec:imp} and illustrated in Section~\ref{sec:app}. Conclusions are drawn in Section~\ref{sec:disc}.  
 
\section{Reaction networks}
\label{sec:stochkin}

Consider a reaction network involving $u$ species $\mathcal{X}_1,
\mathcal{X}_2,\linebreak[1]\ldots,\mathcal{X}_u$ and $v$ reactions $\mathcal{R}_1,\mathcal{R}_2,\ldots,\mathcal{R}_v$ 
such that {reaction $\mathcal{R}_i$ is written as}
\[
 \sum_{j=1}^{u}a_{ij}\mathcal{X}_j \longrightarrow
\sum_{j=1}^{u}b_{ij}\mathcal{X}_j, \quad i=1,\ldots,v 
\]
where $a_{ij}$ denotes the number of molecules of $\mathcal{X}_j$ consumed by reaction $\mathcal{R}_i$ 
and $b_{ij}$ denotes the number of molecules of $\mathcal{X}_j$ produced by reaction $\mathcal{R}_i$. 
Let $X_{j,t}$ denote the (discrete) number of species $\mathcal{X}_j$ at time
$t$, and let $X_t$ be the $u$-vector $X_t = (X_{1,t},X_{2,t},\linebreak[1] \ldots,\linebreak[0] X_{u,t})'$. 
{The effect of a particular reaction is to change the system state $X_t$ abruptly and discretely. Hence, 
if the $i$th reaction occurs at time $t$, the new state becomes
\[
X_t=X_{t-}+S^{i}
\]
where $S^i=(b_{i1}-a_{i1},\ldots,b_{iu}-a_{iu})'$ is the $i$th column of the $u\times v$ stoichiometry matrix $S$. 
The time evolution of $X_t$ is therefore most naturally described by a continuous-time, discrete-valued Markov process 
defined in the following section.}
 
\subsection{Markov jump processes}
\label{sec:mjp}
{We model the time evolution of $X_t$ via a Markov jump process (MJP), so that the state of the system at time $t$ is
\[
X_t=x_0+\sum_{i}S^{i}R_{i,t}
\]
where $x_0$ is the initial system state and $R_{i,t}$ denotes the number of times that the $i$th reaction 
occurs by time $t$. The process $R_{i,t}$ is a counting process with intensity $h_i(x_t)$, known in 
this setting as the reaction hazard, which depends on the current state of the system $x_t$. 
Explicitly, we have that
\[
R_{i,t}=Y_i\left(\int_{0}^{t}h_i(x_s)ds \right)
\]
where the $Y_i$, $i=1,\ldots,v$ are independent, unit rate Poisson processes (see e.g. \cite{kurtz1972} or \cite{Wilkinson06} for further details of 
this representation).} 
The hazard function is given by $h(x_t)=(h_1(x_t),\ldots,\linebreak[1]h_v(x_t))'$. 
Under the standard assumption of mass-action kinetics, $h_i$ is proportional to a product of 
binomial coefficients. That is
\[
h_i(x_t) = c_i\prod_{j=1}^u \binom{x_{j,t}}{a_{ij}}
\] 
where $c_i$ is the rate constant associated with reaction $\mathcal{R}_i$ and $c=(c_1,c_2,\ldots,c_v)'$ 
is a vector of rate constants. {Since in this article,
    except in Section \ref{sec.SIRmodel} the rate constants are assumed to be a known fixed quantities 
we drop them from the notation where possible.} 

Given a value of the initial system state $x_0$, 
exact realisations of the MJP can be generated via \emph{Gillespie's direct method} \citep{Gillespie77}, 
given by Algorithm~\ref{A1}. 

\begin{algorithm}[t]
\caption{Gillespie's direct method}\label{A1}
\begin{enumerate}
\item Set $t=0$. Initialise with $x_0=(x_{1,0},\ldots ,x_{u,0})'$.
\item Calculate $h_{i}(x_t)$, $i=1,\ldots ,v$ and the \emph{combined hazard} $h_{0}(x_t)=\sum_{i=1}^v h_i(x_t)$.
\item Simulate the time to the next event, $t'\sim Exp(h_{0}(x_t))$.
\item Simulate the reaction index, $i$, as a discrete random quantity with
  probability $h_{i}(x_t)/h_{0}(x_t)$, $i=1,\ldots ,v$. 
\item Put $x_{t+t'}:=x_t+S^{i}$, where $S^i$ denotes the $i$th column of $S$.
\item Put $t:=t+t'$. Output $x_t$ and $t$. If $t< T$, return to step~2.
\end{enumerate}
\end{algorithm}

\section{Sampling conditioned MJPs}
\label{sec:cond}

Denote by $\vec{X}=\{X_{s}\,|\, 0< s \leq T\}$ {the MJP sample path 
over the interval $(0,T]$. Complete information on an observed sample 
path $\vec{x}$ corresponds to all reaction times and types. To this end, 
let $n_{r}$ denote the total number of reaction events; reaction times 
(assumed to be in increasing order) and types are denoted by 
$(t_{i},\nu_{i})$, $i=1,\ldots ,n_{r}$, $\nu_{i}\in \{1,\ldots ,v\}$ and we take $t_{0}=0$ and 
$t_{n_{r}+1}=T$.}

Suppose that the initial state $x_{0}$ is a known fixed value and that 
(a subset of components of) the process is observed at time $T$ subject to Gaussian error, 
giving a single observation $y_{T}$ on the random variable
\begin{equation}\label{obs}
Y_{T}=P'x_{T}+\varepsilon_{T}\,,\qquad \varepsilon_{T}\sim \textrm{N}\left(0,\Sigma\right).
\end{equation}   
Here, $Y_{T}$ is a length-$d$ vector, $P$ is a constant matrix of dimension 
$u\times d$ and $\varepsilon_{T}$ is a length-$d$ Gaussian random vector. The role of the 
matrix $P$ is to provide a flexible setup allowing for various observation scenarios. For 
example, taking $P$ to be the $u\times u$ identity matrix corresponds to the case of 
observing all components of $X_t$ (subject to error). We denote the density 
linking $Y_{T}$ and $X_{T}$ as $p(y_{T}|x_{T})$. 

We consider the task of generating trajectories from $p(\vec{x}|x_{0},y_{T})$ given by
\begin{align}
p(\vec{x}|x_{0},y_{T})&= \frac{p(y_{T}|x_{T})p(\vec{x}|x_{0})}{p(y_{T}|x_{0})}\nonumber\\
& \propto p(y_{T}|x_{T})p(\vec{x}|x_{0}) \label{target}
\end{align} 
Here, $p(\vec{x}|x_{0})$ is the {complete data likelihood \citep{Wilkinson06} which takes the form
\[
p(\vec{x}|x_{0})=\left\{\prod_{i=1}^{n_{r}}h_{\nu_{i}}\left(x_{t_{i-1}}\right)\right\}
\exp\left\{-\int_{0}^{T}h_{0}\left(x_t\right)dt\right\}
\] 
where $h_0$ is as defined in line 2 of Algorithm \ref{A1}.} 
Although $p(\vec{x}|x_{0},y_{T})$ will typically be intractable, {generating draws from 
$p(\vec{x}|x_{0})$ is straightforward via Gillespie's direct method (Algorithm~\ref{A1}).} 
This immediately suggests drawing samples from (\ref{target}) using a 
numerical scheme such as weighted resampling. However, as discussed in \cite{GoliWilk15}, 
drawing unconditioned trajectories from $p(\vec{x}|x_{0})$ and weighting by $p(y_{T}|x_{T})$ 
is likely to lead to highly variable weights, unless the level of intrinsic stochasticity 
of $X_t$ is outweighed by the variance of the observation process. {Our umbrella aim, therefore, 
is to find an approximating MJP whose dynamics remain tractable under conditioning on $y_T$. The resulting 
construct can then be used to generate proposed trajectories within the weighted resampling scheme.} 
We will show that this is possible via the derivation of an approximate conditioned hazard function, 
$\tilde{h}(x_t|y_T)$, $t\in(0,T]$, that can be used in place of $h(x_t)$ 
in Algorithm~\ref{A1}. The form for
$\tilde{h}(x_t|y_T)$ that we initially derive depends explicitly on
$t$, so that sampling events might not be straightforward; however
the time-dependence is sufficiently small that it can be ignored and the
resulting bridge mechanism, which has a constant rate between events,
still leads to efficient proposals.

\subsection{Weighted resampling}
\label{sec:wr}

Let $q(\vec{x}|x_{0},y_{T})$ denote the {complete data likelihood for a sample path 
$\vec{x}$ drawn from an} approximate jump process with hazard function $\tilde{h}(x_{t}|y_{T})$. 
The importance weight 
associated with $\vec{x}$ is given by
{\begin{equation}\label{weight0}
w\left(\vec{x}\right)= p(y_{T}|x_{T})\frac{d\mathbb{P}}{d\mathbb{Q}}\left(\vec{x}\right) \nonumber
\end{equation}
where $\frac{d\mathbb{P}}{d\mathbb{Q}}$ is the Radon-Nikodym derivative of the true Markov jump process 
($\mathbb{P}$) with respect to the approximating process ($\mathbb{Q}$) and can be derived in an entirely 
rigorous way \citep{Bremaud81}. An informal approach is provided by \cite{Wilkinson06}, giving the 
Radon-Nikodym derivative as the likelihood ratio
\begin{align}
\frac{d\mathbb{P}}{d\mathbb{Q}}\left(\vec{x}\right) &= p(y_{T}|x_{T})\left\{\prod_{i=1}^{n_{r}}\frac{h_{\nu_{i}}\left(x_{t_{i-1}}\right)}{\tilde{h}_{\nu_{i}}\left(x_{t_{i-1}}|y_{T}\right)}\right\}\exp\left\{-\int_{0}^{T}\left[h_{0}\left(x_t\right)-\tilde{h}_{0}\left(x_t|y_{T}\right)\right]dt\right\} \nonumber
\end{align}
where $h_0(x_t)=\sum_{i=1}^{v}h_i(x_t)$ and $\tilde{h}_{0}(x_t|y_{T})$ is defined analogously. 
As noted above, the explicit dependence of $\tilde{h}$ on $t$ is ignored so that 
both $h_0$ and $\tilde{h}_0$ are piece-wise constant (between reaction events). Hence, in practice, we evaluate the weight using
\begin{align}
 w\left(\vec{x}\right) &= p(y_{T}|x_{T})\left\{\prod_{i=1}^{n_{r}}\frac{h_{\nu_{i}}\left(x_{t_{i-1}}\right)}{\tilde{h}_{\nu_{i}}\left(x_{t_{i-1}}|y_{T}\right)}\right\}
\exp\left\{-\sum_{i=0}^{n_{r}}\left[h_{0}\left(x_{t_{i}}\right)-\tilde{h}_{0}\left(x_{t_{i}}|y_{T}\right)\right]\Delta t_{i}\right\} \label{weight}
\end{align}
where $\Delta t_{i}=t_{i+1}-t_{i}$.} 

The general weighted resampling algorithm is given by Algorithm~\ref{resamp}. 
{It is straightforward to show that the average unnormalised weight gives an unbiased estimator of the transition density $p(y_{T}|x_{0})$. This estimator is given by
\begin{equation}\label{imp0}
\hat{p}(y_{T}|x_0) = \frac{1}{N}\sum_{j=1}^{N}p(y_{T}|X_{T}^j)\frac{p(\vec{X}^{j}|x_{0})}{q(\vec{X}^{j}|x_{0},y_{T})}
\end{equation}
where $\vec{X}^{j}$ is an independent draw from $q(\cdot|x_{0},y_{T})$. In the case of an unknown initial value 
$X_0$ with density $p(x_0)$, Algorithm~\ref{resamp} can be initialised with a sample of size $N$ from $p(x_0)$ in which case 
(\ref{imp0}) can be used to estimate $p(y_T)$.}  
\begin{algorithm}[t]
\caption{Weighted resampling for MJPs}\label{resamp}
\begin{enumerate}
\item For $j=1,2,\ldots ,N$:
\begin{itemize}
\item[(a)] Draw $\vec{x}^j\sim q(\vec{x}|x_{0},y_{T})$ using Algorithm~\ref{A1} with $h(x_t)$ replaced by $\tilde{h}(x_t|y_T)$.
\item[(b)] Construct the unnormalised weight 
\[
\tilde{w}^{j}:=\tilde{w}\left(\vec{x}^j\right)=p(y_{T}|x_{T}^{j})\frac{p(\vec{x}^{j}|x_{0})}{q(\vec{x}^{j}|x_{0},y_{T})}
\]
whose form is given by (\ref{weight}).
\item[(c)] Normalise the weights: $w^{j}=\tilde{w}^{j} / \sum_{i=1}^{N}\tilde{w}^{i}$.
\end{itemize}
\item Resample (with replacement) from the discrete distribution on 
$\big\{\vec{x}^1,\ldots,\vec{x}^N\big\}$ using the normalised weights as probabilities.
\end{enumerate}
\end{algorithm}

It remains for us to find a suitable form of $\tilde{h}(x_{t}|y_{T})$. In what follows, we review two 
existing methods before presenting a novel, alternative approach. Comparisons are made in Section~\ref{sec:app}.

\subsection{Golightly and Wilkinson approach}\label{ch}
The approach of \cite{GoliWilk15} is based on a (linear) Gaussian approximation 
of the number of reaction events in the time between the 
current event time and the next observation time. Suppose 
we have simulated as far as time $t$ and let $\Delta R_{t}$ denote the number of 
reaction events over the time $T-t=\Delta t$. \cite{GoliWilk15} approximate $\Delta R_{t}$ 
by assuming a constant reaction hazard over the whole non-infinitesimal time interval, $\Delta t$. A Gaussian 
approximation to the corresponding Poisson distribution then gives
\[
\Delta R_{t}\sim \textrm{N}\left(h(x_t)\Delta t\,,\,H(x_t)\Delta t\right)
\]
where $H(x_t)=\textrm{diag}\{h(x_t)\}$. Under the Gaussian observation regime 
given by (\ref{obs}) it should be clear that the joint distribution of $\Delta R_{t}$ 
and $Y_{T}$ can then be approximated by
\begin{align*}
 \begin{pmatrix} \Delta R_{t} \\ Y_{T} \end{pmatrix}
&\sim \textrm{N}\left\{\begin{pmatrix} h(x_t)\Delta t \\ P'\left(x_{t}+S\,h(x_t)\Delta t\right)\end{pmatrix}\,,\, \right. \left.\begin{pmatrix} H(x_t)\Delta t & H(x_t)S'P\Delta t\\
P'S\,H(x_t)\Delta t & P'S\,H(x_t)S'P\Delta t +\Sigma\end{pmatrix}\right\}.
\end{align*}
Taking the expectation of $(\Delta R_{t}|Y_{T}=y_{T})$ and dividing by $\Delta t$ gives 
an approximate conditioned hazard as
\begin{align}
&\tilde{h}(x_t|y_{T})=h(x_t)\nonumber\\
&\qquad +H(x_t)S'P\left(P'S\,H(x_t)S'P\Delta t +\Sigma\right)^{-1}\left(y_{T}-P'\left[x_{t}+S\,h(x_t)\Delta t\right]\right). \label{haz}
\end{align}
By ignoring the explicit time dependence of $\tilde{h}(x_t|y_{T})$
(\emph{i.e.}, after each most-recent event, until the next event, fixing
$\Delta t$ to its value at the most recent event), we can use (\ref{haz}), suitably 
truncated to ensure positivity, in 
Algorithm~\ref{A1} to give trajectories $\vec{x}^i$, $i=1,\ldots,N$, to be used in Algorithm~\ref{resamp}. 
Whilst use of (\ref{haz}) has been shown to work well in several applications, assumptions of normality 
of $\Delta R_t$ and that the hazard is constant over a time interval of length $\Delta t$ are often unreasonable, as 
we will show.  

\subsection{Fearnhead approach}\label{fh}

As noted by \cite{fearnhead2008} (see also \cite{ruttor09}), an expression for the intractable conditioned hazard 
can be derived exactly. Consider again an interval $[0,T]$ and suppose that we have simulated as far as time 
$t\in[0,T]$. For reaction $\mathcal{R}_i$ let $x'=x_{t}+S^{i}$. Recall that $S^{i}$ denotes the $i$th column of the stoichiometry 
matrix so that $x'$ is the state of the MJP after a single occurrence of $\mathcal{R}_i$. The conditioned 
hazard of $\mathcal{R}_i$ satisfies
\begin{align}
h_{i}(x_t|y_T)&=\lim_{\delta t\to 0}\frac{Pr(X_{t+\delta t}=x'|X_{t}=x_{t},y_{T})}{\delta t} \nonumber \\
&=h_{i}(x_t)\lim_{\delta t\to 0}\frac{p(y_{T}|X_{t+\delta t}=x')}{p(y_{T}|X_{t}=x_t)} \nonumber \\
&=h_{i}(x_t)\frac{p(y_{T}|X_{t}=x')}{p(y_{T}|X_{t}=x_t)}. \label{fearn}
\end{align}
In practice, the intractable transition density $p(y_{T}|x_{t})$ must be replaced by a suitable 
approximation. \cite{GoliKyp17} (see also \cite{fearnhead2008} for the case of no measurement error) 
used the transition density governing the (discretised) chemical Langevin 
equation (CLE). The CLE \citep{Gillespie92b,Gillespie00} is an It\^o stochastic differential equation (SDE) 
that has the same infinitesimal mean and variance as the MJP. It is written as
\begin{equation}\label{cle}
dX_t = S\,h(X_t)dt + \sqrt{S\operatorname{diag}\{h(X_t)\}S'}\,dW_t,
\end{equation}
where $W_t$ is a $u$-vector of standard Brownian motion and $\sqrt{S\operatorname{diag}\{h(X_t)\}S'}$ 
is a $u\times u$ matrix $B$ such that $BB'=S\operatorname{diag}\{h(X_t)\}S'$. Since the CLE can rarely 
be solved analytically, it is common to work with a discretisation such as the Euler-Maruyama discretisation:
\begin{equation}\label{CLEdisc}
X_{t+\delta t}-X_{t} = S\,h(X_t)\delta t + \sqrt{S\operatorname{diag}\{h(X_t)\}S'\delta t}\,Z
\end{equation}
where $Z$ is a standard multivariate Gaussian random variable. Combining (\ref{CLEdisc}) with the observation model 
(\ref{obs}) gives an approximate conditioned hazard as
\begin{equation}\label{fearnCLE}
\tilde{h}_{i}(x_t|y_T)=h_{i}(x_t)\frac{p_{\textrm{cle}}(y_{T}|X_{t}=x')}{p_{\textrm{cle}}(y_{T}|X_{t}=x_t)}
\end{equation}
where
\begin{align*}
p_{\textrm{cle}}(y_{T}|X_{t}=x_t)&=N\left(y_T; P'(x_t+S\,h(x_t)\Delta t)\,,\right. \left. \,P'S\,H(x_t)S'P\Delta t+\Sigma\right)
\end{align*}
with $p_{\textrm{cle}}(y_{T}|X_{t}=x')$ defined similarly. As with the approach of \cite{GoliWilk15}, the remaining 
time $\Delta t$ until the observation is treated as a single discretisation. However, unless $\Delta t=T-t$ is very small, 
$p_{\textrm{cle}}$ is unlikely to achieve a reasonable approximation of the transition probability under the jump process. In 
what follows, therefore, we seek an approximation that is both accurate and computationally inexpensive.

\section{Improved constructs}
\label{sec:imp}

We take (\ref{fearn}) as a starting point and replace $p(y_{T}|X_{t}=x')$ and $p(y_{T}|X_{t}=x_t)$ 
using the linear noise approximation (LNA)
\citep{kurtz1970,elf2003,Komorowski09,schnoerr17}. We first describe the LNA, and
then consider two constructions for bridges from a known initial
condition, $x_0$, to a potentially noisy observation $Y_T$, based on different implementations 
of the LNA. The first is expected to be more accurate as the
approximate hazard is recalculated after every event by re-integrating a set
of ODEs from the event time to the observation time both from the
current value and once for each
possible next reaction. The second is more
computationally efficient as the recalculation is based on a single, initial
integration of a set of ODEs from time $0$ to time $T$.

\subsection{Linear noise approximation}

For notational simplicity we rewrite the CLE in (\ref{cle}) as
\begin{equation}\label{sde}
dX_t = \alpha(X_t)dt + \sqrt{\beta(X_t)}\,dW_t 
\end{equation}
where
\[
\alpha(X_t)= S\,h(X_t),\qquad \beta(X_t)=S\,\textrm{diag}\{h(X_t)\}S'
\]
and derive the LNA by directly approximating (\ref{sde}). The basic idea behind 
construction of the LNA is to adopt the partition $X_t=z_t+M_t$ where 
the deterministic process $z_t$ satisfies an ordinary differential equation
\begin{equation}\label{mean}
\frac{dz_{t}}{dt}=\alpha(z_{t})
\end{equation} 
and the residual stochastic process $M_t$ can be well approximated under the 
assumption that residual stochastic fluctuations are ``small'' relative to 
the deterministic process. Taking the first two terms in the Taylor expansion of $\alpha(X_t)$, 
and the first term in the Taylor expansion of $\beta(X_t)$ gives an 
SDE satisfied by an approximate residual process $\tilde{M}_t$ of the form
\begin{equation}\label{residual}
d\tilde{M}_t = F_t\tilde{M}_t\,dt + \sqrt{\beta(z_t)}\,dW_t,
\end{equation}
where $F_t$ is the Jacobian matrix with $(i,j)$th element 
$(F_t)_{i,j}=\partial\alpha_i(z_t)/\partial z_{j,t}$. The SDE in (\ref{residual}) 
can be solved by first defining the $u\times u$ fundamental matrix $G_t$ as the solution 
of 
\begin{equation}\label{fund}
\frac{dG_{t}}{dt}=F_tG_t,\qquad G_0=I_u,
\end{equation}
where $I_u$ is the $u\times u$ identity matrix. Under the assumption of 
a fixed or Gaussian initial condition, $\tilde{M}_0\sim N(m_{0},V_{0})$, 
it can be shown that \citep[see e.g.][]{fearnhead14}
\[
\tilde{M}_t|\tilde{M}_0=m_0\sim N(G_t m_0,G_t\psi_t G_t')
\]
where $\psi_t$ satisfies
\begin{equation}\label{psi}
\frac{d\psi_{t}}{dt}=G_t^{-1}\beta(z_t,c)\left(G_t^{-1}\right)'.
\end{equation}
It is convenient here to write $V_t=G_t\psi_t G_t'$ and it is straightforward to 
show that $V_t$ satisfies
\begin{equation}\label{var}
\frac{dV_{t}}{dt}=V_t F_t'+\beta(z_t,c)+F_t V_t.
\end{equation}
In practice, if $x_0$ is a known fixed value then we may take $z_0=x_0$, $m_0=0_u$ 
(the $u$-vector of zeros) and $V_0=0_{u\times u}$ (the $u\times u$ zero matrix). 
Solving (\ref{mean}) and (\ref{var}) gives the approximating distribution of $X_t$ as 
\[
X_t|X_0=x_0 \sim N(z_t,V_t).
\]
In this case, the ODE system governing the fundamental matrix $G_t$ need not be solved. 

\subsection{LNA bridge with restart}
\label{sec:lnar}

Now, consider again the problem of approximating the MJP transition probability $p(y_{T}|X_{t}=x_t)$. 
Given a value $x_t$ at time $t\in[0,T)$, the ODE system given by 
(\ref{mean}) and (\ref{var}) can be re-integrated over the time interval $(t,T]$ to give output 
denoted by $z_{T|t}$ and $V_{T|t}$. Similarly, the initial conditions are denoted $z_{t|t}=x_t$ 
and $V_{t|t}=0_{u\times u}$. We refer to use of the LNA 
in this way as the LNA with restart (LNAR). The approximation to $p(y_{T}|X_{t}=x_t)$ is given 
by
\[
p_{\textrm{lnar}}(y_{T}|X_{t}=x_t)=N\left(y_T; P'z_{T|t}\,,\,P'V_{T|t}P+\Sigma\right).
\]  
Likewise, $p_{\textrm{lnar}}(y_{T}|X_{t}=x')$ can be obtained by initialising (\ref{mean}) 
with $z_{t|t}=x'$ and integrating again. 
Hence, the approximate conditioned hazard is given by
\begin{equation}\label{fearnLNAR}
\tilde{h}_{i}(x_t|y_T)=h_{i}(x_t)\frac{p_{\textrm{lnar}}(y_{T}|X_{t}=x')}{p_{\textrm{lnar}}(y_{T}|X_{t}=x_t)}
\end{equation}
Whilst use of the LNA in this way is likely to give an accurate approximation to the intractable 
transition probability (especially as $t$ approaches $T$), the conditioned hazard in (\ref{fearn}) 
must be calculated for $x_t$ and for each $x'$ obtained after the $v$ possible transitions of the process. 
Consequently, the ODE system given by (\ref{mean}) and (\ref{var})
must be solved at each event time for each of the $v+1$ possible 
states. Since the LNA ODEs are rarely tractable (necessitating the use of a numerical 
solver), this approach is likely to be 
prohibitively expensive, computationally. In the next section, we outline a novel strategy for reducing the 
cost associated with integrating the LNA ODE system, that only requires one full integration. 

\subsection{LNA bridge without restart}
\label{sec:lna}

Consider the solution of the ODE system given by (\ref{mean}), (\ref{fund}) and (\ref{psi}) 
over the interval $(0,T]$ with respective initial conditions $Z_0=x_0$, $G_0=I_u$ and $\psi_0=0_{u\times u}$. 
Although in practice a numerical solver must be used, we assume that the solution can be obtained 
over a sufficiently fine time grid to allow reasonable approximation to the ODE solution at an arbitrary time $t\in (0,T]$, denoted 
by $z_t$, $G_t$ and $\psi_t$.

Given a value $x_t$ at time $t\in[0,T)$, the LNA (without restart) approximates the intractable transition probability under the MJP by
\begin{align*}
p_{\textrm{lna}}(y_{T}|X_{t}=x_t)&=N\left(y_T; P'[z_{T}+G_{T|t}(x_t-z_t)]\,,\right. \left. \,P'[G_{T|t}\psi_{T|t}G_{T|t}']P+\Sigma\right)
\end{align*}
where $G_{T|t}$ and $\psi_{T|t}$ are the solutions of (\ref{fund}) and (\ref{psi}) integrated over $(t,T]$ with initial conditions 
$G_{t|t}=I_{u}$ and $\psi_{t|t}=0_{u\times u}$. Crucially, the ODE system satisfied by $z_t$ is not re-integrated (and hence the residual 
term at time $t$ is $\tilde{M}_t=x_t-z_t$). Moreover, $G_{T|t}$ and $\psi_{T|t}$ can be obtained without further integration. We have 
that
\begin{align*}
\textrm{E}(\tilde{M}_T|\tilde{M}_0 = m_0)&=G_T m_0 \\
&= G_{T|t}\textrm{E}(\tilde{M}_t|\tilde{M}_0 = m_0)\\
&=G_{T|t}G_t m_0
\end{align*}
and therefore the first identity we require is
\begin{equation}\label{idG}
G_{T|t}=G_{T}G_{t}^{-1}.
\end{equation}
Similarly,
\begin{align*}
\textrm{Var}(\tilde{M}_T|\tilde{M}_0=m_0) &= G_T \psi_T G_T\\
&=G_{T|t}\textrm{Var}(\tilde{M}_t|\tilde{M}_0=m_0)G_{T|t}'\\
&\quad +G_{T|t}\psi_{T|t}G_{T|t}'\\
&=G_T\psi_t G_T' + G_{T|t}\psi_{T|t}G_{T|t}'\\
&=G_T \psi_t G_T' +G_T G_t^{-1}\psi_{T|t}(G_t')^{-1}G_T'
\end{align*}
where we have used (\ref{idG}) to obtain the last line. The second identity we require is therefore
\begin{equation}\label{idPsi}
\psi_{T|t}=G_{t}(\psi_{T}-\psi_{t})G_{t}'.
\end{equation}
Hence, given $z_t$, $G_t$ and $\psi_t$ for $t\in (0,T]$, $p_{\textrm{lna}}(y_{T}|X_{t}=x_t,c)$ is easily 
evaluated via repeated application of (\ref{idG}) and (\ref{idPsi}). 
Additionally obtaining $p_{\textrm{lna}}(y_{T}|X_{t}=x')$ is straightforward by replacing the residual $x_t-z_t$ with $x'-z_t$. 
Hence, only one full integration of (\ref{mean}), (\ref{fund}) and (\ref{psi}) over $(0,T]$ is required, giving a computationally 
efficient construct. The conditioned hazard takes the form
\begin{equation}\label{fearnLNA}
\tilde{h}_{i}(x_t,c|y_T)=h_{i}(x_t)\frac{p_{\textrm{lna}}(y_{T}|X_{t}=x')}{p_{\textrm{lna}}(y_{T}|X_{t}=x_t)}
\end{equation} 
In Section \ref{sec:disc} we describe how, in the case of unknown $X_0$ it is
possible to make further computational savings, using this technique.

The accuracy of $p_{\textrm{lna}}$ (and therefore the accuracy of the resulting 
conditioned hazard) is likely to depend on $T$, the length of the inter-observation period over which a realisation of the conditioned process 
is required. For example, the residual process $\tilde{M}_t$ 
will approximate the true (intractable) residual process increasingly poorly if 
$z_t$ and $X_t$ diverge significantly as $t$ increases. We investigate the effect of inter-observation time in the next section.

\section{Applications}
\label{sec:app}

In order to examine the empirical performance of the methods proposed 
in section~\ref{sec:imp}, we consider three examples of increasing complexity. These are a 
simple (and tractable) death model, the stochastic Lotka-Volterra 
model examined by \citet{BWK08} among others and a
susceptible-infected-removed (SIR) epidemic 
model. For the last of these, we use the best performing LNA-based construct to drive a pseudo-marginal 
Metropolis-Hastings (PMMH) scheme to perform fully Bayesian inference for the rate constants $c$. 
Using real data consisting of susceptibles and infectives during 
the well studied Eyam plague \citep{Raggett82}, we compare bridge-based PMMH with a standard implementation 
(using blind, forward simulation) and a recently proposed scheme based on the alive particle filter 
\citep{drovandi15}. All algorithms are coded in R and were run on a desktop computer with an Intel Core 
i7-4770 processor at 3.40GHz. 

\subsection{Death model}

We consider a single reaction, governing a single specie $\mathcal{X}$, of the form
\[
\mathcal{R}_{1}:\,\mathcal{X} \longrightarrow \emptyset
\]
with associated hazard function  
\[
h(x_t) = c\,x_{t}
\]
where $x_t$ denotes the state of the system at time $t$. 

Under the assumption of an error free observation scenario, 
the conditioned hazard of \cite{GoliWilk15}, given by (\ref{haz}), 
takes the form
\[
\tilde{h}(x_t|y_{T})=\frac{x_T-x_t}{\Delta t}
\] 
and recall that $\Delta t=T-t$. 

The CLE is given by
\[
dX_{t}=-c\,X_{t}\,dt + \sqrt{c\,X_{t}}\,dW_{t}.
\]
Although the CLE is tractable in this special case \citep{Cox85}, for reaction networks of reasonable size and complexity, the CLE will be intractable. 
We therefore implement the approach of \cite{fearnhead2008} by taking the conditioned 
hazard as in (\ref{fearnCLE}) where $p_{\textrm{cle}}$ is based on a single-time step numerical approximation of the CLE. 
The Euler-Maruyama approximation gives 
\[
p_{\textrm{cle}}(x_T|x_t)=N\left(x_t\,;\, x_{t}-c\,x_{t}\,\Delta t\,,\,c\,x_{t}\,\Delta t\right).
\] 

The ODE system characterising the LNA (equations (\ref{mean}), (\ref{fund}) and (\ref{psi})) with respective 
initial conditions $z_0=x_0$, $G_0=I_u$ and $\psi_0=0_{u\times u}$ can be solved analytically to give
\[
z_t=x_0e^{-c\, t},\qquad G_t=e^{-c\, t},\qquad \psi_t= x_0 \left(e^{c\, t}-1\right).
\]
Hence, for the LNA with restart, we have that
\[
p_{\textrm{lnar}}(x_{T}|x_t)=N\left\{x_T\,;\, x_t e^{-c\, \Delta t}\,,\,x_t e^{-c\, \Delta t}\left(1-e^{-c\, \Delta t}\right)  \right\}.
\] 
For the LNA without restart, we obtain
\[
p_{\textrm{lna}}(x_{T}|x_t)=N\left\{x_T\,;\, x_t e^{-c\, \Delta t}  \,,\,x_0 e^{-c\, T}\left(1-e^{-c\, \Delta t}\right)  \right\}.
\]

In what follows, we took $c=0.5$ and $x_{0}=50$ to be fixed. The end-point $x_T$ 
was chosen as either the median, lower 1\% or upper 99\% quantile of the forward 
process $X_{T}|X_{0}=50$. We adopt the notation that $x_{T,(\alpha)}$ 
is the $\alpha\%$ quantile of $X_{T}|X_{0}=50$. Hence, we took the end-point 
$x_T \in\{x_{T,(1)},x_{T,(50)},x_{T,(99)}\}$. To assess the performance of the proposed approach 
as an observation is made with increasing time sparsity, we took $T\in\{0.5,1,2\}$. We applied weighted resampling 
(Algorithm~\ref{resamp}) with five different hazard functions. These were, the unconditioned `blind' hazard function, 
the conditioned hazard of Golightly/Wilkinson given by (\ref{haz}), and the Fearnhead approach based on the CLE (\ref{fearnCLE}), LNA with restart 
(\ref{fearnLNAR}) and LNA without restart (\ref{fearnLNA}). The resulting algorithms are designated as \emph{blind}, \emph{GW}, \emph{F-CLE}, 
\emph{F-LNAR} and \emph{F-LNA}. Each was run $m=5000$ times with $N=10$ samples to give a set of 5000 estimates of the transition probability 
$\pi(x_{t}|x_{0})$ and we denote this set by $\widehat{\pi}^{1:m}(x_{t}|x_{0})$. 
To compare the algorithms, we report the effective sample size 
\[
\textrm{ESS}(\widehat{\pi}^{1:m}) = \frac{\left(\sum_{i=1}^{m}\widehat{\pi}^{i}\right)^{2}}{\sum_{i=1}^{m}\left(\widehat{\pi}^{i}\right)^{2}}
\]
and relative mean-squared error $\textrm{ReMSE}(\widehat{\pi}^{1:m})$ given by
\[
\textrm{ReMSE}(\widehat{\pi}^{1:m})=\frac{1}{m}\sum_{i=1}^{m}
\frac{\left[\widehat{\pi}^{i}(x_{t}|x_{0})-\pi(x_{t}|x_{0})\right]^{2}}{\pi(x_{t}|x_{0})}
\]
where $\pi(x_{t}|x_{0})$ can be obtained analytically (e.g., \cite{bailey1964}) as
\[
\pi(x_{t}|x_{0})=\begin{pmatrix} x_0 \\ x_t \end{pmatrix}
e^{-c\,t\, x_t}\left(1-e^{-c\,t}\right)^{x_0-x_t}.
\]

The results are summarised in Table~\ref{tab:tabD}. Whilst the Blind approach gives broadly comparable performance 
to the conditioned approaches when $x_T=x_{T,(50)}$, its performance deteriorates significantly when the end-point 
is taken to be a value in the tails of $x_{t}|X_{0}=50$. This is due to the Blind approach struggling to generate 
trajectories that are highly unlikely to hit the neighbourhood
  of the end-point. For the CH approach we see a decrease in ESS and an increase 
in ReMSE as $T$ increases, due to the linear form being unable to adequately describe the exponential like decay 
exhibited by the true conditioned process. Whilst the F-CLE approach performs well when $x_T=x_{T,(50)}$ and 
$x_T=x_{T,(99)}$, it is unable to match the performance of the LNA-based methods across all scenarios. The effect of not 
restarting the LNA (i.e. by reintegrating the LNA ODEs after each value of the jump process is generated) appears 
to be minimal here, with both F-LNAR and F-LNA giving comparable ESS and ReMSE values.

\begin{table*}[t]
\centering
\small
	\begin{tabular}{@{}lccccc@{}}
         \toprule
\phantom{Algorithm}  & Blind & CH & F-CLE & F-LNAR & F-LNA  \\
\midrule
\multicolumn{6}{@{}l}{$X_T=x_{T,(50)}$}\\
\quad $T=0.5$&3026, 8.9$\times 10^{-2}$ &4142, 2.8$\times 10^{-2}$ &3782, 4.4$\times 10^{-2}$ &3852, 4.1$\times 10^{-2}$ &3751, 4.5$\times 10^{-2}$   \\
\quad $T=1$  &2806, 8.8$\times 10^{-2}$ &3528, 4.8$\times 10^{-2}$ &3594, 4.5$\times 10^{-2}$ &3856, 3.3$\times 10^{-2}$ &3648, 4.3$\times 10^{-2}$  \\
\quad $T=2$  &1540, 8.2$\times 10^{-2}$ &1161, 3.9$\times 10^{-1}$ &3564, 4.7$\times 10^{-2}$ &3966, 3.0$\times 10^{-2}$ &3900, 3.3$\times 10^{-2}$  \\
\midrule
\multicolumn{6}{@{}l}{$X_T=x_{T,(1)}$}\\
\quad $T=0.5$&\phantom{0}247, 1.1$\times 10^{-1}$ &3969, 1.8$\times 10^{-3}$           &3228, 2.5$\times 10^{-3}$ &3278, 2.5$\times 10^{-3}$ &3107, 2.9$\times 10^{-3}$  \\
\quad $T=1$  &\phantom{0}339, 1.1$\times 10^{-1}$ &3194, 3.8$\times 10^{-3}$           &2106, 9.3$\times 10^{-3}$ &3515, 2.9$\times 10^{-3}$ &3281, 3.6$\times 10^{-3}$  \\
\quad $T=2$  &\phantom{00}73, 3.1$\times 10^{-2}$ &\phantom{0}135, 1.9$\times 10^{-1}$ &1015, 2.1$\times 10^{-2}$ &3275, 2.6$\times 10^{-3}$ &2894, 3.7$\times 10^{-3}$  \\
\midrule
\multicolumn{6}{@{}l}{$X_T=x_{T,(99)}$}\\
\quad $T=0.5$&\phantom{0}646, 1.0$\times 10^{-1}$ &4316, 2.3$\times 10^{-3}$ &3926, 3.8$\times 10^{-3}$ &4042, 3.5$\times 10^{-3}$ &3995, 3.7$\times 10^{-3}$  \\
\quad $T=1$  &\phantom{0}436, 9.9$\times 10^{-2}$ &3901, 2.6$\times 10^{-3}$ &3806, 2.8$\times 10^{-3}$ &4017, 2.3$\times 10^{-3}$ &3938, 2.5$\times 10^{-3}$  \\
\quad $T=2$  &\phantom{0}223, 5.2$\times 10^{-2}$ &1660, 2.1$\times 10^{-2}$ &3660, 3.7$\times 10^{-3}$ &4067, 2.3$\times 10^{-3}$ &3862, 3.0$\times 10^{-3}$  \\
\bottomrule
\end{tabular}
      \caption{Death model. $\textrm{ESS}(\widehat{\pi}^{1:m})$ and 
$\textrm{ReMSE}(\widehat{\pi}^{1:m})$, based on 5000 runs of each algorithm.}\label{tab:tabD}	
\end{table*}

\subsection{Lotka-Volterra}\label{app:lv}

We consider here a Lotka-Volterra model of prey ($\mathcal{X}_1$) and predator ($\mathcal{X}_2$) interaction comprising 
three reactions of the form
\begin{align*}
\mathcal{R}_1:\quad \mathcal{X}_{1} &\xrightarrow{\phantom{a}c_{1}\phantom{a}} 2\mathcal{X}_{1}\\
\mathcal{R}_2:\quad \mathcal{X}_{1}+\mathcal{X}_{2} &\xrightarrow{\phantom{a}c_{2}\phantom{a}} 2\mathcal{X}_{2}\\
\mathcal{R}_3:\quad \mathcal{X}_{2} &\xrightarrow{\phantom{a}c_{3}\phantom{a}} \emptyset.
\end{align*}
The stoichiometry matrix is given by
\[
S = \left(\begin{array}{rrr} 
1 & -1 & 0\\
0 & 1 & -1
\end{array}\right)
\]
and the associated hazard function is 
\[
h(x_t) = (c_{1}x_{1,t}, c_{2}x_{1,t}x_{2,t}, c_{3}x_{2,t})'.
\]
The conditioned hazard described in Section~\ref{ch} and given by (\ref{haz}) can then be obtained. 

The CLE for the Lotka-Volterra model is given by
\begin{align}
 d\begin{pmatrix}X_{1}\\ X_{2}\end{pmatrix}&= \begin{pmatrix}
				c_{1}X_{1}-c_{2}X_{1}X_{2} \\
				c_{2}X_{1}X_{2}-c_{3}X_{2}
			\end{pmatrix}\,dt + \begin{pmatrix}c_{1}X_{1}+c_{2}X_{1}X_{2} & -c_{2}X_{1}X_{2} \\
			 -c_{2}X_{1}X_{2}	 & c_{2}X_{1}X_{2}+c_{3}X_{2}
			\end{pmatrix}^{1/2}\,d \begin{pmatrix}W_{1}\\ W_{2}\end{pmatrix} \label{lv}
\end{align}
after suppressing dependence on $t$. It is then straightforward to obtain the Euler-Maruyama approximation of the CLE, for use 
in the conditioned hazard described in Section~\ref{fh} and given by (\ref{fearnCLE}).

For the linear noise approximation, the Jacobian 
matrix $F_t$ is given by
\[
F_{t} = \left(\begin{array}{cc} 
c_{1}-c_{2}z_{2,t} & -c_{2}z_{1,t} \\
c_{2}z_{2,t} & c_{2}z_{1,t}-c_{3} 
\end{array}\right).
\]
Unfortunately, the ODEs characterising the LNA solution, given by (\ref{mean}), (\ref{fund}) and (\ref{psi}) are intractable, 
necessitating the use of a numerical solver. In what follows, we use the \texttt{deSolve} package in R, with the default lsoda 
integrator \citep{petzold83}. 

Our initial experiments used the following settings. Following \cite{BWK08} among others we imposed the parameter values 
$c=(c_1,c_2,c_3)'=(0.5,0.0025,\linebreak[1]0.3)'$ and let $x_0=(50,50)'$. 
We assumed an observation model of the form (\ref{obs}) and took $\Sigma=\sigma^2 I_{2}$ 
with $\sigma=5$ representing low measurement error (since typical simulations of $X_{1,t}$ and 
$X_{2,t}$ are around two orders of magnitude larger than $\sigma$). We generated a number of challenging scenarios 
by taking $y_T$ as the pair of 1\%, 50\% or 99\% marginal quantiles of $Y_T|X_0=(50,50)'$ for $T\in\{1,2,3,4\}$. These quantiles 
are denoted by $y_{T,(1)}$, $y_{T,(50)}$ and $y_{T,(99)}$ respectively, and are shown in Table~\ref{tab:tabLV}. 

\begin{table*}[t]
\centering
\small
	\begin{tabular}{@{}lcccc@{}}
         \toprule
   & $T=1$ & $T=2$ & $T=3$ & $T=4$ \\
\midrule
  $y_{T,(1)}$ & (53.34, 27.99) & (75.83, 22.59) & (109.51, 20.90) & (157.34, 23.65) \\
  $y_{T,(50)}$& (73.25, 58.43) & (108.69, 39.92)& (162.03, 41.23) & (238.62, 49.89) \\ 
  $y_{T,(99)}$& (95.33, 58.43) & (147.28, 58.26)& (225.77, 64.19) & (337.65, 83.79) \\
\bottomrule
\end{tabular}
      \caption{Lotka-Volterra model. Quantiles of $Y_T|X_0=(50,50)'$ found by repeatedly simulating from the Euler-Maruyama 
approximation of (\ref{lv}) with $c=(0.5,0.0025,0.3)'$ and corrupting $X_{1,T}$ and $X_{2,T}$ with additive $N(0,5^2)$ noise.}\label{tab:tabLV}	
\end{table*}

\begin{figure*}[t]
\centering
\psfrag{CH}[][][1.3][0]{CH}
\psfrag{F-LNA}[][][1.3][0]{F-LNA}
\psfrag{F-LNAR}[][][1.3][0]{F-LNAR}
\psfrag{F-CLE}[][][1.3][0]{F-CLE}
\psfrag{Blind}[][][1.3][0]{Blind}
\psfrag{Xt}[][][1.5][0]{$X_t$}
\psfrag{t}[][][1.5][0]{$t$}
\includegraphics[angle=270,width=\textwidth]{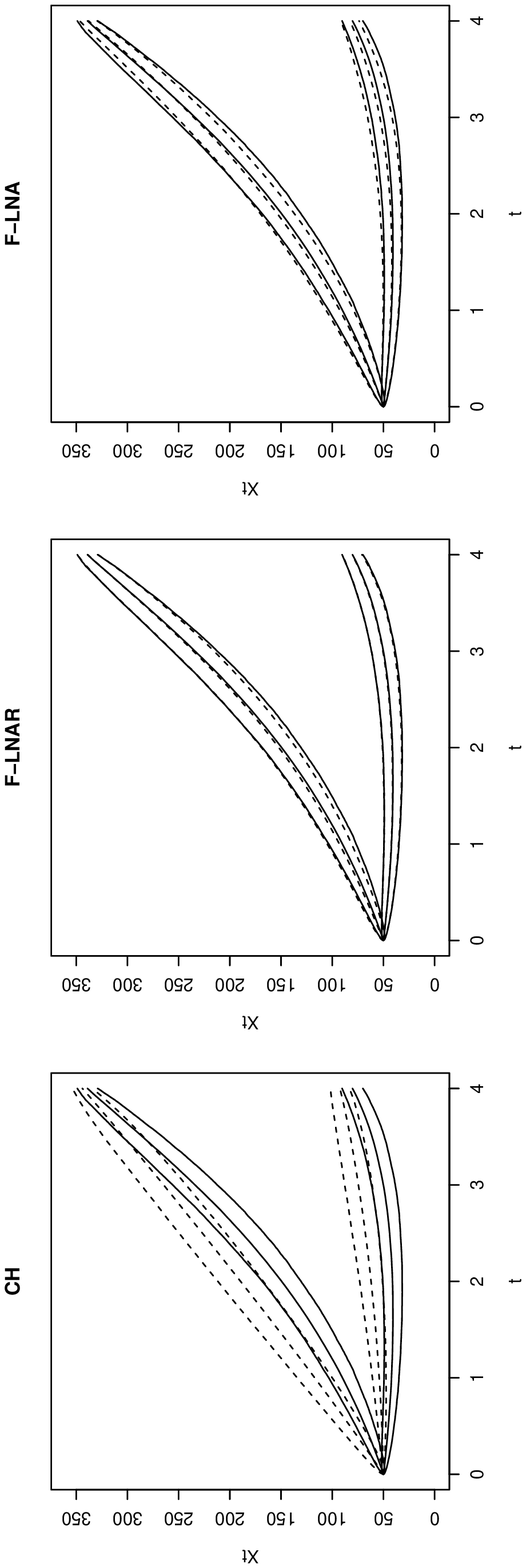}
\includegraphics[angle=270,width=\textwidth]{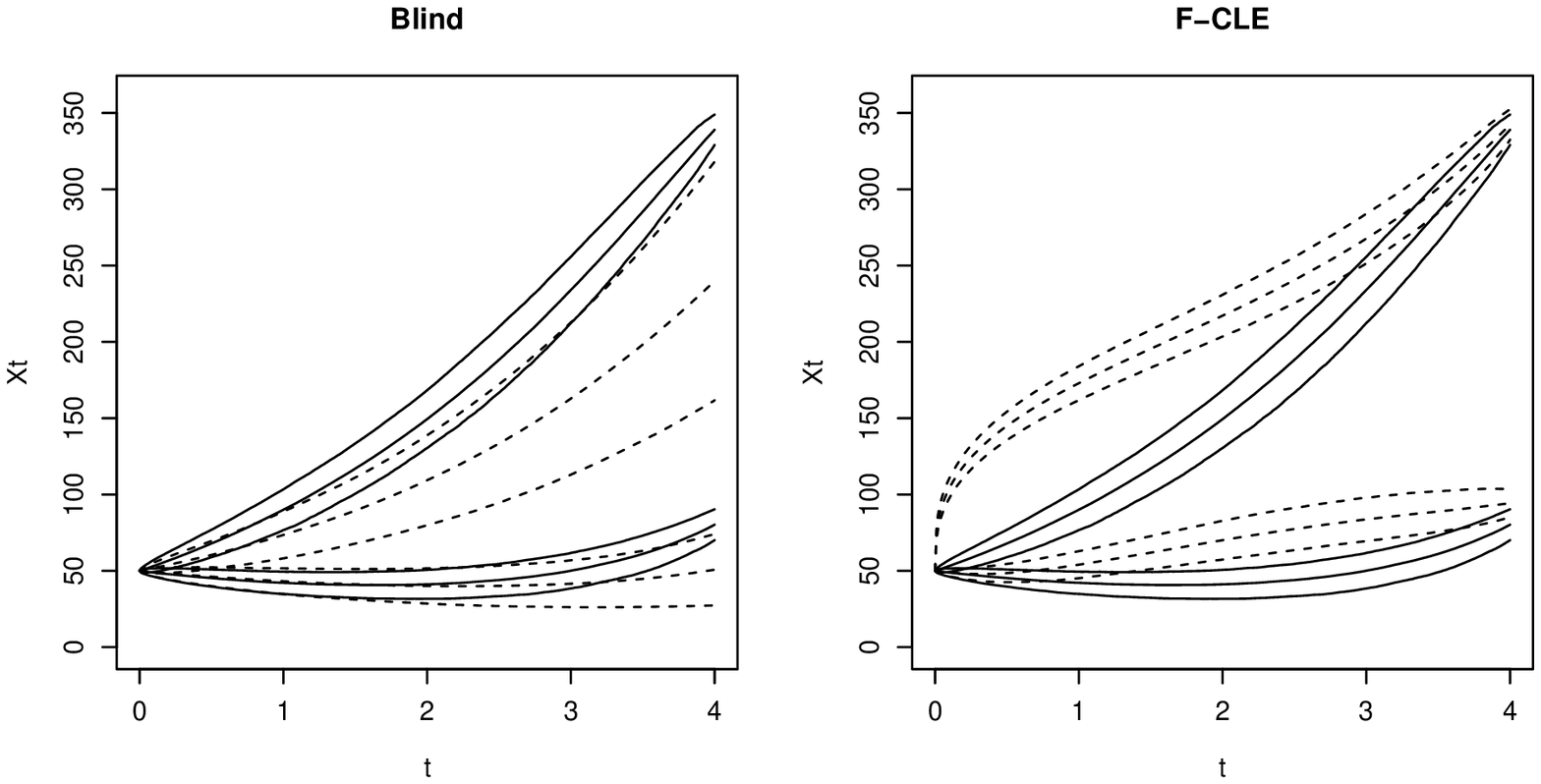}
\caption{Lotka-Volterra model. Mean and two standard deviation intervals for the true conditioned 
process $X_{t}|x_0,y_T$ (solid lines) and various bridge constructs (dashed lines) using $y_T=y_{T,(99)}$, $T=4$ and $\sigma=5$. The upper lines correspond to 
the prey component and the lower lines correspond to the predator component.}\label{fig:figLV1}
\end{figure*}

Figure~\ref{fig:figLV1} compares summaries (mean plus and minus two standard deviations) of each competing bridge process with the same summaries 
of the true conditioned process (obtained via simulation), for the extreme case of $T=4$ and $y_T=y_{T,(99)}$. Plainly, the blind forward simulation approach 
and CLE-based Fearnhead approach (F-CLE) are unable to match the dynamics of the true conditioned process. Moreover, 
we found that these bridges gave very small effective sample sizes for $T\geq 2$ and we therefore omit these results from the following analysis. 

\begin{figure*}[t]
\centering
\psfrag{(a)}[][][1.3][0]{$y_{T,(1)}$}
\psfrag{(b)}[][][1.3][0]{$y_{T,(50)}$}
\psfrag{(c)}[][][1.3][0]{$y_{T,(99)}$}
\psfrag{T}[][][1.5][5]{$T$}
\psfrag{ESS}[][][1.5][0]{ESS}
\psfrag{logESSs}[][][1.5][0]{$\log_{10}(\textrm{ESS/s})$}
\psfrag{logCPU}[][][1.5][0]{$\log_{10}(\textrm{CPU})$}
\includegraphics[angle=270,width=\textwidth]{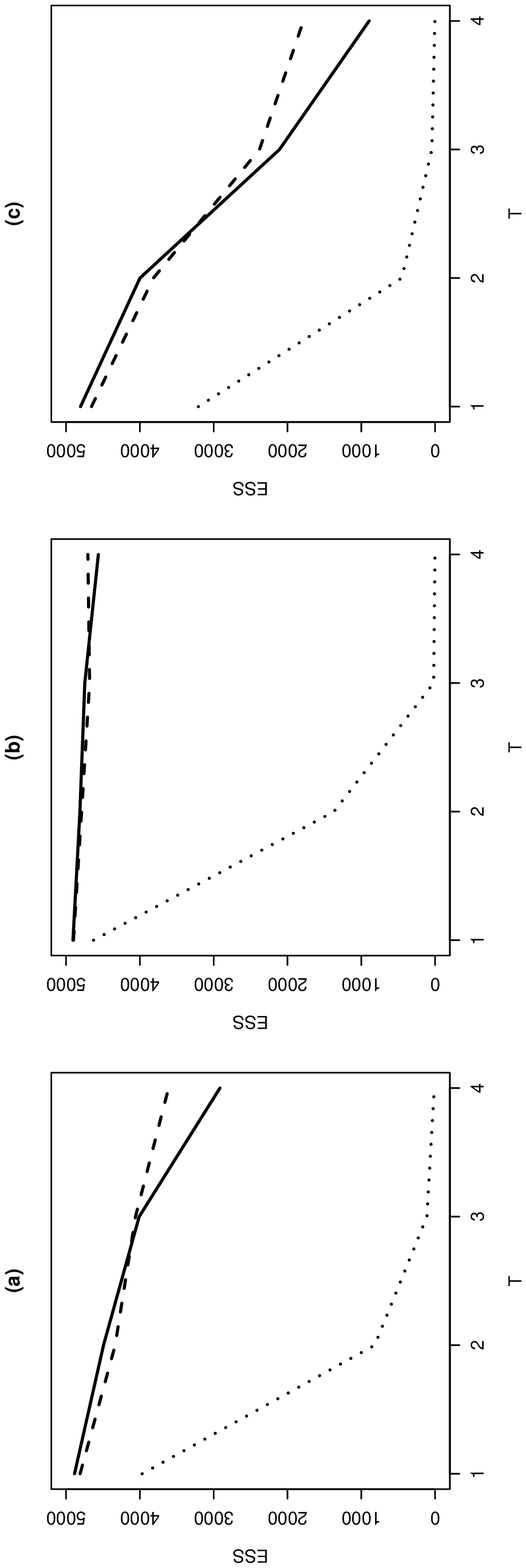}
\includegraphics[angle=270,width=\textwidth]{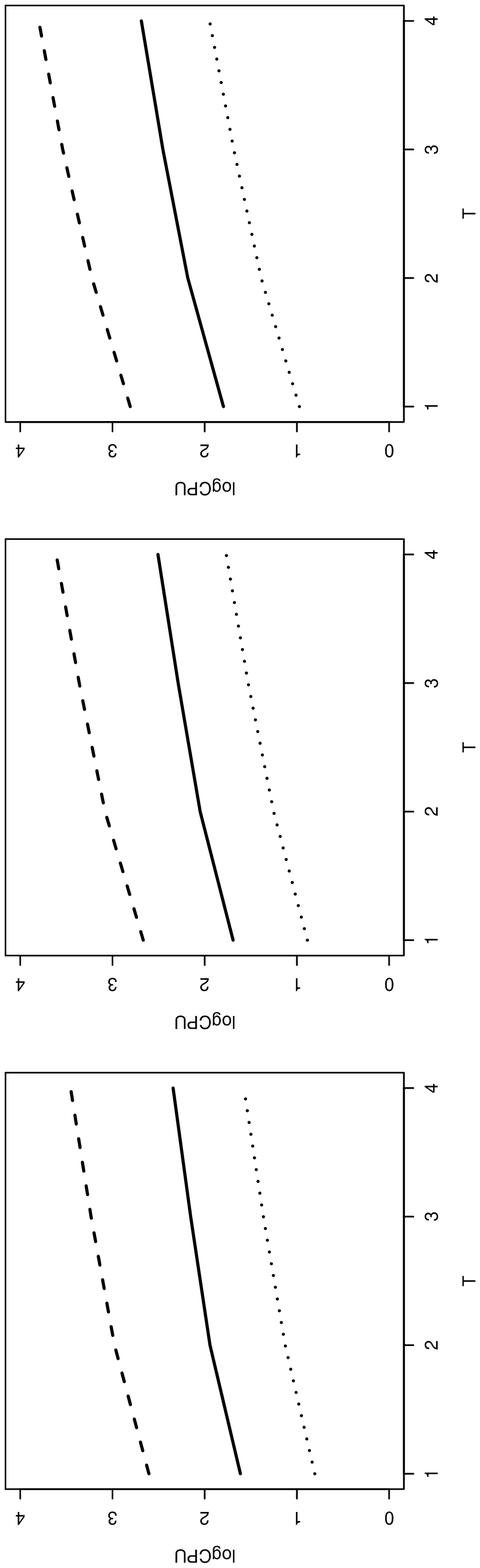}
\includegraphics[angle=270,width=\textwidth]{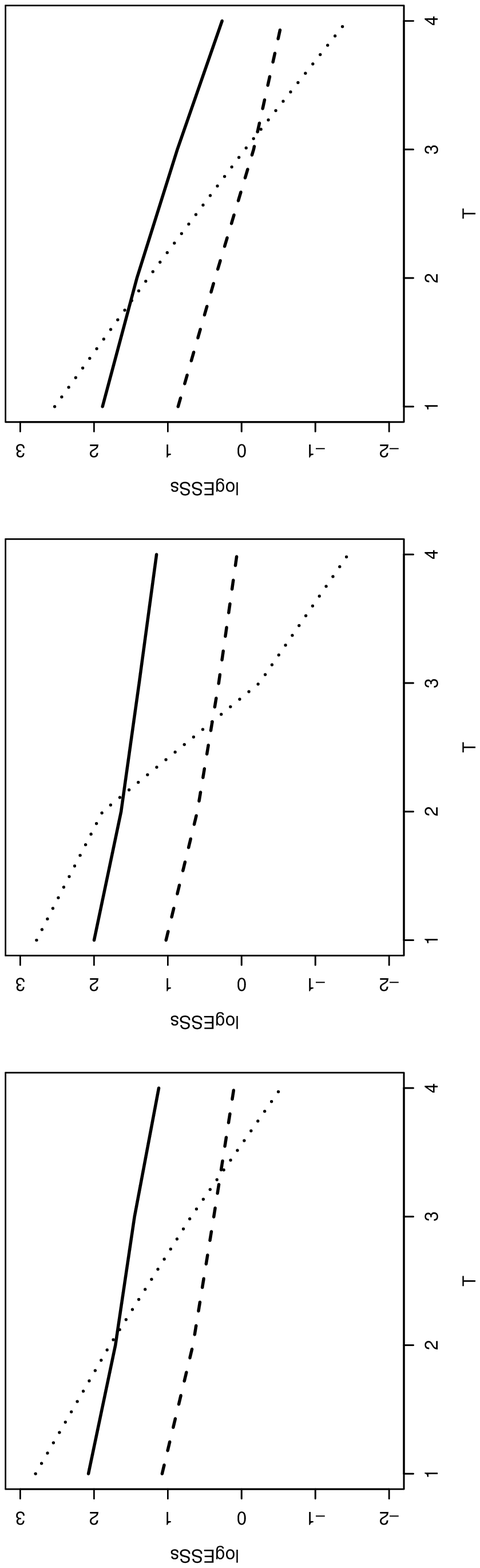}
\caption{Lotka-Volterra model. Effective sample size (ESS, top row), log (base 10) computing time in seconds (CPU, middle row) and log (base 10) effective sample 
size per second (ESS/s, bottom row) based on the output of the weighted resampling algorithm with $N=5000$ and $y_T\in\{y_{T,(1)},y_{T,(50)},y_{T,(99)}\}$, $T=1,2,3,4$. 
Dotted lines: CH. Dashed lines: F-LNAR. Solid lines: F-LNA.}\label{fig:figLV2}
\end{figure*}

We report results based on weighted resampling using $N=5000$ with three different hazard functions: the Golightly/Wilkinson approach (CH) and the Fearnhead approach based on the 
LNA with and without restart (F-LNAR and F-LNA respectively). For the latter (F-LNA), we integrated the LNA once in total. Figure~\ref{fig:figLV2} shows, for each value of $y_T$ in Table~\ref{tab:tabLV}, 
effective sample size (ESS), log (base 10) CPU time and log (base 10) ESS per second. Note that for this example, ESS is calculated as
\[
\textrm{ESS}(\tilde{w}^{1:N}) = \frac{\left(\sum_{i=1}^{N}\tilde{w}^{i}\right)^{2}}{\sum_{i=1}^{N}\left(\tilde{w}^{i}\right)^{2}}
\]  
where $\tilde{w}^{1:N}$ denotes the unnormalised weights generated by the weighted resampling algorithm. We see that 
although CH is computationally inexpensive, ESS decreases as $T$ increases, as it is unable to match the nonlinear dynamics of the 
true conditioned process. In contrast, although more computationally expensive, F-LNAR and F-LNA maintain high ESS values as $T$ 
is increased. Consequently, in terms of ESS per second, CH is outperformed by F-LNAR for $T\geq 3$ and F-LNA for 
$T\geq 2$. Due to not having to restart the LNA ODEs after each simulated value of the jump process, F-LNA is around an 
order of magnitude faster than F-LNAR in terms of CPU time, with the difference increasing as $T$ is increased. Given then 
the comparable ESS values obtained for F-LNAR and F-LNA, we see that in terms of ESS/s, F-LNA outperforms F-LNAR by at least 
an order of magnitude in all cases, and outperforms CH by 1-2 orders of magnitude when $T=4$.  

\begin{figure*}[t]
\centering
\psfrag{(a)}[][][1.3][0]{}
\psfrag{(b)}[][][1.3][0]{}
\psfrag{(c)}[][][1.3][0]{}
\psfrag{T}[][][1.5][5]{$T$}
\psfrag{ESS}[][][1.5][0]{ESS}
\psfrag{logESSs}[][][1.5][0]{$\log_{10}(\textrm{ESS/s})$}
\psfrag{logCPU}[][][1.5][0]{$\log_{10}(\textrm{CPU})$}
\includegraphics[angle=270,width=\textwidth]{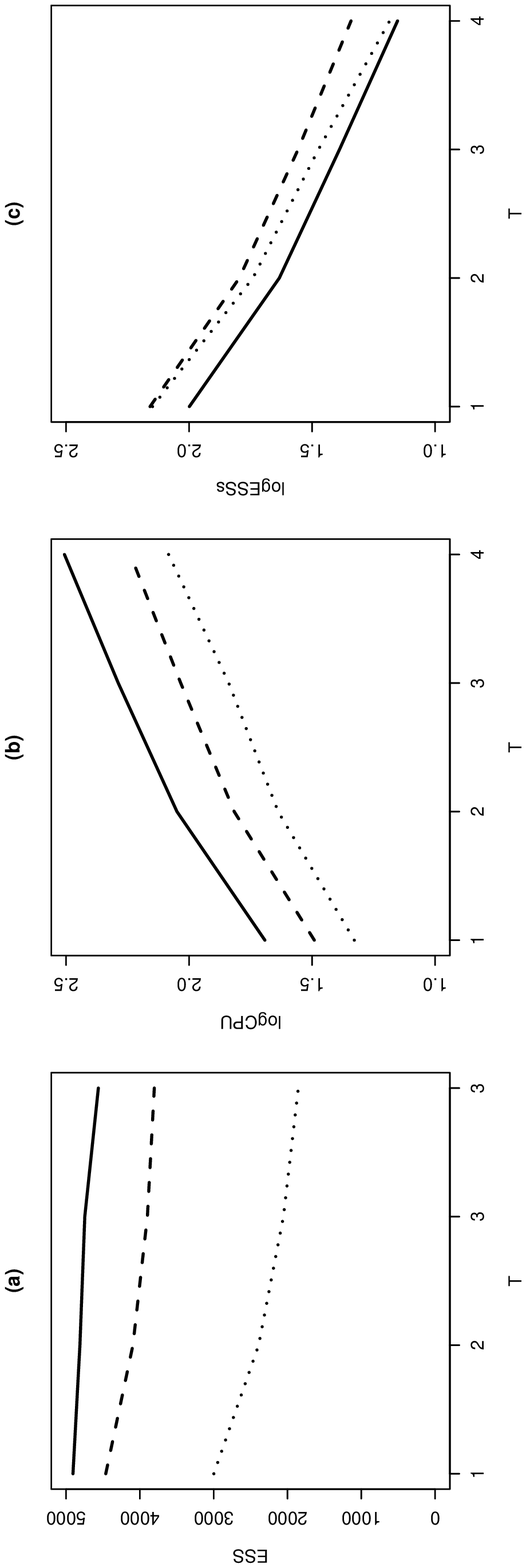}
\caption{Lotka-Volterra model. (a) Effective sample size (ESS, left panel), log (base 10) computing time in seconds (CPU, middle panel) and log (base 10) effective sample 
size per second (ESS/s, right panel) based on the output of the weighted resampling algorithm with $N=5000$ and $y_T=y_{T,(50)}$, $T=1,2,3,4$. 
Dotted lines: $x_0=(10,10)'$ and $\sigma=1$. Dashed lines: $x_0=(25,25)'$ and $\sigma=2.5$. Solid lines: $x_0=(50,50)'$ and $\sigma=5$.}\label{fig:figLVR1}
\end{figure*}

The LNA is known to break down as an inferential model in situations involving low counts of the MJP components \citep{schnoerr17}. 
Therefore, to investigate the performance of the use of the LNA in constructing an approximate conditioned hazard in low count scenarios, we additionally 
considered an initial condition with 
$x_{1,0}=x_{2,0}\in\{10,25,50\}$ and took $y_T$ as the median of $Y_T|X_0=x_0$ for $T\in\{1,2,3,4\}$. To fix the 
relative effect of the measurement error, we took $\sigma=1$ for the case $x_0=(10,10)'$ and scaled $\sigma$ 
in proportion to the components of $x_0$ for the remaining scenarios. The resulting values of $y_T$ can be found in Table~\ref{tab:tabLVR1}. 
We report results based on weighted resampling using $N=5000$ and F-LNA in Figure~\ref{fig:figLVR1}. We see that 
when the initial condition is decreased from $x_0=(50,50)'$ to $x_0=(10,10)'$, 
ESS decreases by a factor of around 1.6 (4906 vs 2998) when $T=1$ and 2.5 (4562 vs 1853) when $T=4$. Nevertheless, 
computational cost decreases as $x_0$ decreases (and in turn, the expected number of reaction events in the observation 
window decreases). Hence, there is little difference in overall efficiency (ESS/s) across the three scenarios. 

\begin{table*}[t]
\centering
\small
	\begin{tabular}{@{}lccccc@{}}
         \toprule
$x_0'$ & $\sigma$  & $T=1$ & $T=2$ & $T=3$ & $T=4$ \\
\midrule
$(10,10)$ & $1$  & (15.80, 7.68) & (25.46, 5.94) & (41.17, 4.72) & (67.11, 3.92) \\
$(25,25)$ & $2.5$& (38.67, 20.04) & (60.72, 16.71)& (96.09, 14.92) & (152.50, 14.87) \\ 
$(50,50)$ & $5$  & (73.25, 58.43) & (108.69, 39.92)& (162.03, 41.23) & (238.62, 49.89) \\ 
\bottomrule
\end{tabular}
      \caption{Lotka-Volterra model. Median of $Y_T|X_0=x_0$ found by repeatedly simulating from the Euler-Maruyama 
approximation of (\ref{lv}) with $c=(0.5,0.0025,0.3)'$ and corrupting $X_{1,T}$ and $X_{2,T}$ with additive $N(0,\sigma^2)$ noise.}\label{tab:tabLVR1}	
\end{table*}

\subsection{SIR model}
\label{sec.SIRmodel}
\subsubsection{Model and data}
The Susceptible--Infected--Removed (SIR) epidemic model has two species 
(susceptibles $\mathcal{X}_{1}$ and infectives $\mathcal{X}_{2}$) and two reaction channels (infection of 
a susceptible and removal of an infective):
\begin{align*}
\mathcal{R}_1:\quad \mathcal{X}_{1}+\mathcal{X}_{2} &\xrightarrow{\phantom{a}c_1\phantom{a}} 2\mathcal{X}_{2}\\
\mathcal{R}_2:\quad \mathcal{X}_{2} &\xrightarrow{\phantom{a}c_2\phantom{a}} \emptyset.
\end{align*}
The vector of rate constants is $c=(c_1,c_2)'$ and the stoichiometry matrix is given by
\[
S = \left(\begin{array}{rr} 
-1 & 0\\
 1 & -1
\end{array}\right).
\]
The hazard function is given by $h(x_t) = (c_1 x_{1,t}x_{2,t},\linebreak[1] c_2 x_{2,t})'$. 
For the linear noise approximation, the Jacobian matrix $F_t$ is given by
\[
F_{t} = \left(\begin{array}{cc} 
-c_1 z_{2} & -c_1 z_{1} \\
c_1 z_{2} & c_1 z_{1}-c_2 
\end{array}\right).
\]  
The ODEs characterising the LNA solution, given by (\ref{mean}), (\ref{fund}) and (\ref{psi}) are intractable. As 
in Section~\ref{app:lv}, we use the \texttt{deSolve} package in R whenever a numerical solution is required. 

We consider data consisting of 8 observations on susceptible and infectives during the outbreak of plague in the village of Eyam, England. 
The data are taken over a four month period from June 18th 1666 and are presented here in Table~\ref{tab:tabE}. Note that the infective population 
is estimated from a list of deaths, and by assuming a fixed illness length \cite[]{Raggett82}.

\begin{table*}[t]
  \centering
  \small
  \begin{tabular}{@{} lllllllll@{}}
\toprule
& \multicolumn{8}{@{}c}{Time (months)}\\
                 &0  &0.5  &1  &1.5  &2   &2.5  &3  &4 \\
\midrule    
    Susceptibles &254  &235  &201  &153  &121   &110  &97  &83 \\
    Infectives   &\phantom{00}7  &\phantom{0}14  &\phantom{0}22  &\phantom{0}29  &\phantom{0}20   &\phantom{00}8  &\phantom{0}8  &\phantom{0}0  \\ 
\bottomrule
  \end{tabular}
  \caption{Eyam plague data.}\label{tab:tabE}
\end{table*}

\subsubsection{Pseudo-marginal Metropolis-Hastings}
Let $\vec{y}=\{y_{t_i}\}$, $i=1,\ldots,8$ denote the observations at times $0=t_1 <\ldots < t_8 = 4$. 
The latent Markov jump process over the time interval $(t_i,t_{i+1}]$ is denoted by $\vec{X}_{(t_i,t_{i+1}]}=\{X_{s}\,|\, t_i< s \leq t_{i+1}\}$. 
Under the assumption of no measurement error, we have that $X_{t_i}=y_{t_{i}}$, $i=1,\ldots,8$. 
Upon ascribing a prior density $p(c)$ to the rate constants $c$, Bayesian inference may proceed 
via the marginal parameter posterior
\begin{equation}
p(c|\vec{y})\propto p(c)p(\vec{y}|c)
\end{equation}
where
\begin{equation}\label{obslike}
p(\vec{y}|c)=\prod_{i=1}^{7}p(y_{t_{i+1}}|y_{t_{i}},c)
\end{equation} 
is the observed data likelihood. Although $p(\vec{y}|c)$ is intractable, 
we note that each term in (\ref{obslike}) can be seen as the normalising constant 
of 
\[
p(\vec{x}_{(t_i,t_{i+1}]}|x_{t_i},y_{t_{i+1}},c)\propto p(y_{t_{i+1}}|x_{t_{i+1}})p(\vec{x}_{(t_i,t_{i+1}]}|x_{t_i},c)
\]
where $p(y_{t_{i+1}}|x_{t_{i+1}})$ takes the value 1 if $x_{t_{i+1}}=y_{t_{i+1}}$ and 0 otherwise. Hence, running steps 
1(a) and (b) of Algorithm~\ref{resamp} with $x_0$ and $y_T$ replaced by $x_{t_i}$ and $y_{t_{i+1}}$ respectively, 
can be used to unbiasedly estimate $p(y_{t_{i+1}}|y_{t_{i}},c)$. No resampling is required, since only those trajectories 
that coincide with the observation $y_{t_{i+1}}$ will have non-zero weight. By analogy with equation (\ref{imp0}), and allowing explicit dependence 
on $c$ we have the unbiased estimator
\begin{align}
\hat{p}(y_{t_{i+1}}|y_{t_{i}},c) &= \frac{1}{N}\sum_{j=1}^{N}p(y_{t_{i+1}}|X_{t_{i+1}}^j)\frac{p(\vec{X}_{(t_i,t_{i+1}]}^{j}|x_{t_i},c)}{q(\vec{X}_{(t_i,t_{i+1}]}^{j}|x_{t_i},y_{t_{i+1}},c)}\label{imp}
\end{align}
where $\vec{X}_{(t_i,t_{i+1}]}^{j}$ is an independent draw from \sloppy$q(\cdot|x_{t_i},y_{t_{i+1}},c)$. Then, 
multiplying the $\hat{p}(y_{t_{i+1}}|y_{t_{i}},c)$, $i=1,\ldots,7$, gives an unbiased estimator of the observed data likelihood $p(\vec{y}|c)$. 

An alternative unbiased estimator of the observed data likelihood can be found by using (a special case of) the alive particle filter \citep{delmoral2015}. Essentially, 
forward draws are repeatedly generated from $p(\cdot|x_{t_i},c)$ (via Gillespie's direct method) until $N+1$ trajectories that match the observation are obtained. Let 
$n_i$ denote the number of simulations required to generate $N+1$ matches with $y_{t_{i+1}}$. The estimator is then given by
\begin{equation}\label{alive}
\hat{p}(y_{t_{i+1}}|y_{t_{i}},c)= \frac{N}{n_i - 1}.
\end{equation} 

Let $U\sim p(\cdot|c)$ denote the flattened vector of all random variables required to generate the estimator of observed data likelihood, which we denote by 
$\hat{p}_{U}(\vec{y}|c)$. The pseudo-marginal Metropolis-Hastings (PMMH) scheme is an MH scheme that targets the joint density
\[
p(c,u)\propto p(c)\hat{p}_{u}(\vec{y}|c)p(u|c)
\]
for which it is easily checked that 
\begin{align*}
\int p(c,u)\, du &\propto p(c)\int \hat{p}_{u}(\vec{y}|c)p(u|c)\,du\\
& \propto p(c)p(\vec{y}|c)
\end{align*}
where the last line follows from the unbiasedness property of $\hat{p}_{U}(\vec{y}|c)$. Hence we see that the target posterior 
$p(c|\vec{y})$ is a marginal of the joint density $p(c,u)$. Now, running an MH scheme with a proposal density of the form $q(c^*|c)p(u^*|c^*)$ gives the 
acceptance probability
\[
\textrm{min}\left\{1, \frac{p(c^*)\hat{p}_{u^*}(\vec{y}|c^*)}{p(c)\hat{p}_{u}(\vec{y}|c)}\times \frac{q(c|c^*)}{q(c^*|c)}\right\}.
\]  
Practical advice for 
choosing $N$ to balance mixing performance and computational cost can found in 
\cite{doucet15} and \cite{sherlock2015}. The variance of the log-posterior 
(denoted $\sigma^{2}_{N}$, computed with $N$ samples) at a central value of $c$ 
(e.g. the estimated posterior median) should be around 2. In what follows, we 
use a random walk on $\log c$ as the parameter proposal. The innovation variance 
is taken to be the marginal posterior variance of $\log c$ estimated from a pilot run, 
and further scaled to give an acceptance rate of around 0.2--0.3. We followed 
\cite{Ho2018} by adopting independent $N(0,100^2)$ priors for $\log c_i$, $i=1,2$.

Although we do not pursue it here, the case of non-zero measurement error is easily 
accommodated by iteratively running Algorithm~\ref{resamp} in full, for each observation 
time $t_i$, $i=1,\ldots,7$. At time $t_i$, $y_T$ is replaced by 
$y_{t_{i+1}}$ and $x_0$ is replaced by $x_{t_i}^j$. At time $t_1$, $x_0$ can be replaced 
by a draw from a prior density $p(x_{t_1})$ placed on the unobserved initial value. The product (across time) of the average 
unnormalised weight can be shown to give an unbiased estimator of the observed 
data likelihood \citep{delmoral04,pitt12}. We refer the reader to \cite{GoliWilk15} and the references therein for further details 
of the resulting Metropolis-Hastings scheme.

\subsubsection{Results}

We ran PMMH using the observed data likelihood estimator based on (\ref{imp}), with trajectories drawn 
either using forward simulation or the Fearnhead approach based on the LNA (without restart). We designate the 
former as ``Blind'' and the latter as ``F-LNA''. Additionally, we ran PMMH using the observed data likelihood 
estimator based on (\ref{alive}). We designate this scheme as ``Alive''.

\begin{figure*}[t]
\centering
\psfrag{c1}[][][1.5][0]{$c_1$}
\psfrag{c2}[][][1.5][0]{$c_2$}
\includegraphics[angle=270,width=\textwidth]{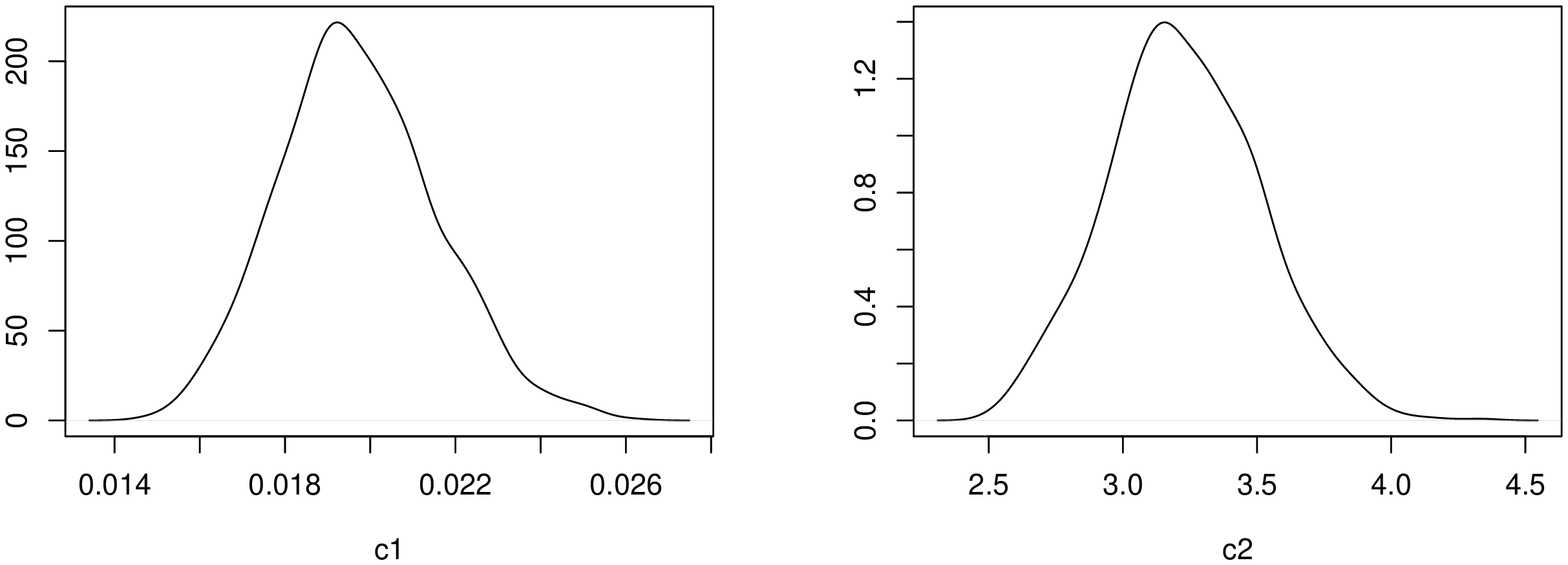}
\caption{SIR model. Marginal posterior densities based on the output of F-LNA.}\label{fig:figSIR2}
\end{figure*}

\begin{figure*}[t]
\centering
\psfrag{It}[][][1.5][0]{$I_t$}
\psfrag{St}[][][1.5][0]{$S_t$}
\psfrag{t}[][][1.5][0]{$t$}
\includegraphics[angle=270,width=\textwidth]{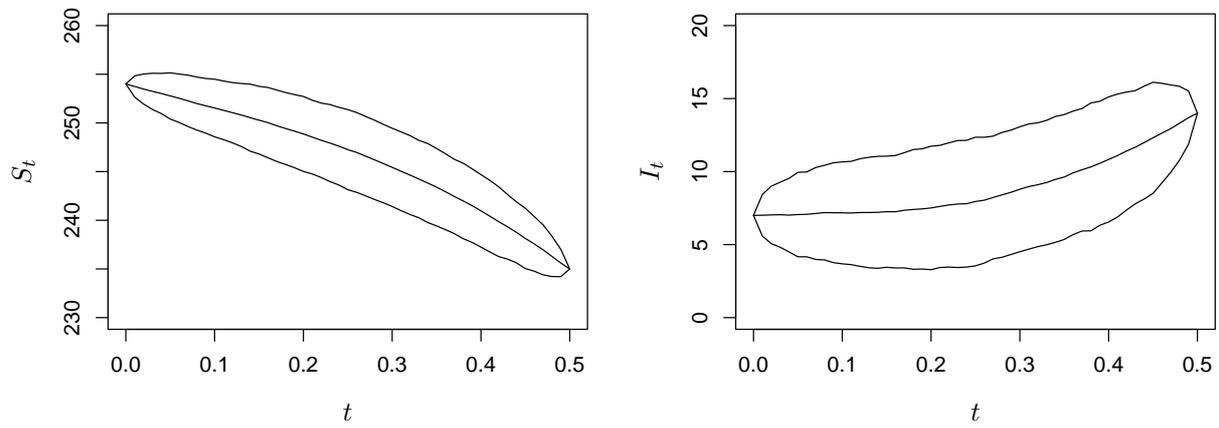}
\caption{SIR model. Mean and two standard deviation intervals for the true conditioned 
process $X_{t}|x_0,y_{0.5},c$ over the first observation interval, with $c=(0.02,3.2)'$.}\label{fig:figSIR1}
\end{figure*}

\begin{table*}[t]
  \centering
  \small
  \begin{tabular}{@{}llllll@{}}
    \toprule
    Algorithm  &  $N$ & CPU (s) & mESS & mESS/s & Rel.  \\
    \midrule
    Alive       & \phantom{000}8  &126697            &737  &0.0058 & 1\phantom{.0}\\
    Blind       & 5000            &\phantom{0}68177  &863  &0.0127 & 2.2\\
    F-LNA       &  \phantom{0}100 &\phantom{0}25752  &644  &0.0250 & 4.3\\
    \bottomrule
  \end{tabular}
  \caption{SIR model. Number of particles $N$, 
    CPU time (in seconds $s$), minimum ESS, minimum ESS per second and relative (to Blind) minimum ESS per second. All 
    results are based on $10^4$ iterations of each scheme.}\label{tab:tabE2}	
\end{table*}

We ran each scheme for $10^4$ iterations. For Alive, we followed \cite{drovandi16} by terminating any likelihood calculation that exceeded 
100,000 forward simulations, and rejecting the corresponding move. Marginal posterior densities can be found in Figure~\ref{fig:figSIR2} and are 
consistent with the posterior summaries reported by \cite{Ho2018}. Figure~\ref{fig:figSIR1} summarises the 
posterior distribution of $X_{t}|x_0,y_{0.5},c$, where $c$ is fixed at the estimated posterior mean. We note the nonlinear 
behaviour of the conditioned process over this time interval, with similar nonlinear dynamics observed for other intervals 
(not reported). Table~\ref{tab:tabE2} summarises the computational and statistical performance of the competing inference schemes. 
We measure statistical efficiency by calculating minimum (over each parameter chain) effective sample size per second (mESS/s). 
As is appropriate for MCMC output, we use
\[
\textrm{ESS}=\frac{n_{\textrm{iters}}}{1+2\sum_{k=1}^{\infty}\alpha_k}
\]
where $\alpha_k$ is the autocorrelation function for the series at lag $k$ and $n_{\textrm{iters}}$ is the number 
of iterations in the main monitoring run. Inspection of Table~\ref{tab:tabE2} reveals that although use of the alive particle 
filter only requires $N=8$ (compared to $N=5000$ and $N=100$ for Blind and F-LNA respectively), it exhibits the largest 
CPU time. We found that for parameter values in the tails of the posterior, Alive would often require many thousands 
of forward simulations to obtain $N=8$ matches. Consequently Alive is outperformed by Blind by a factor of 2 in terms of 
overall efficiency. Use of the LNA-driven bridge (without restart) gives a further improvement over Blind of a factor of 2. 

\section{Discussion}
\label{sec:disc}

Performing efficient sampling of a Markov jump process (MJP) between a
known value and a potentially partial or noisy observation  
is a key requirement of simulation-based approaches to parameter inference. Generating end-point conditioned 
trajectories, known as bridges, is challenging due to the intractability of the probability function governing 
the conditioned process. Approximating the hazard function associated with the conditioned process (that is, the 
conditioned hazard), and correcting draws obtained via this hazard function using weighted 
resampling or Markov chain Monte Carlo offers a viable 
solution to the problem. Recent approaches in this direction \citep{fearnhead2008,GoliWilk15} give approximate 
hazard functions that utilise a Gaussian approximation of the MJP. For example, \cite{GoliWilk15} 
approximates the number of reactions between observation times as Gaussian. \cite{fearnhead2008} 
recognises that the conditioned hazard can be written in terms of the intractable transition 
probability associated with the MJP. The transition probability is replaced with a Gaussian transition density 
obtained from the Euler-Maruyama approximation of the chemical Langevin equation. In both approaches the 
remaining time until the next observation is treated as a single discretisation. Consequently, 
the accuracy of the resulting bridges deteriorates as the inter-observation time increases. 

Starting with the form of the conditioned hazard function, we have proposed a novel bridge 
construct by replacing the intractable MJP transition probability with the transition density 
governing the linear noise approximation (LNA). Whilst our approach also involves a Gaussian approximation, 
we find that the tractability of the LNA can be exploited to give an accurate bridge construct. Essentially, 
the LNA solution can be re-integrated over each observation window to maintain accuracy. The cost of `restarting' 
the LNA in this way is likely to preclude its practical use. We have therefore further proposed an implementation 
that only requires a single full integration of the ordinary differential equation system governing the LNA. Our experiments demonstrated superior performance of the 
LNA based bridge over existing constructs, especially in data-sparse scenarios. Whilst the LNA is known 
to give a poor approximation of the MJP in low count scenarios \citep{schnoerr17}, we note that its role here is in the 
approximation of transition densities over ever diminishing time intervals. Moreover, the 
resulting approximate conditioned hazard function is corrected for via a weighted resampling scheme. 
Consequently, we find that use of the LNA in this way is relatively robust to situations involving low 
counts. Using a real data application, we further demonstrated the potential 
of the proposed methodology in allowing efficient parameter inference. 

When the dimension of the statespace is finite then the
  transition probability from a known state at time $0$ to a known
  state at time $T$, can be calculated exactly and efficiently via the
  action of a matrix exponential on a vector
  \cite[e.g.][]{SidjeStewart1999}, giving the likelihood directly; alternatively
the uniformisation method of \cite{rao13} may be used for Bayesian
inference. The recent article \cite{georgoulas17} extends the
standard finite-statespace matrix-exponential method to an infinite
statespace pseudo-marginal MCMC algorithm which uses random truncation
\cite[e.g.][]{GlynnRhee2014} to produce a realisation from an unbiased estimator of
the likelihood when the observations are exact. In contrast to the algorithms which we have
investigated, which simulate paths for the process and whose performance 
improves as the observation noise increases, any extension to the
algorithm of \cite{georgoulas17} that allows for observation error
would reduce the efficiency of the algorithm. This suggestes the
possibility that for small enough observation noise an extension to the algorithm in
\cite{georgoulas17} might be more efficient than our non-restarting
bridge. Investigations in to the relative efficiencies of such
algorithms are ongoing.

This article has focused on bridges from a known initial
condition. When the initial condition is unknown, such as typically
arises in a particle filter-based analysis, a sample from the distribution of the initial state,
$\{x_0^{1},\ldots,x_0^N\}$, is available
and a separate bridge to the observation is required from each element
of the sample. In this case, two different implementations of the LNA
bridge without restarting are possible.  In the first implementation, trajectories 
$\vec{X}^i|x_0^i,y_T$ are generated using one full integration of (\ref{mean}), (\ref{fund}) and (\ref{psi}) over $(0,T]$ 
\emph{for each} $x_0^i$. That is, each trajectory has (\ref{mean}) initialised at $x_0^i$. In the second implementation, 
(\ref{mean}), (\ref{fund}) and (\ref{psi}) are integrated \emph{just once}, irrespective of the number of required trajectories. 
This can be achieved by initialising (\ref{mean}) at some plausible value e.g. $E(X_0)$. Although the second implementation will 
be more computationally efficient than the first, some loss of
accuracy is expected, especially when the uncertainty in $X_0$ is
large. A single integral, however, may well be
 adequate in the cases which are the focus of this article: where the
observation noise is small. Investigating the efficiency of the bridge construct 
in this scenario, as well as in multi-scale settings \citep[see e.g.][]{thomas14} where some reactions regularly occur more frequently than others, remains the 
subject of ongoing research.

\bibliographystyle{apalike}
\bibliography{bridgebib}

\appendix

\end{document}